\documentclass[]{spie}  

\usepackage{amsmath,amsfonts,amssymb}
\usepackage{graphicx}
\usepackage[colorlinks=true, allcolors=blue]{hyperref}
\usepackage{epstopdf}
\usepackage{subfig,acronym,upgreek}

\title{Modal noise characterisation of a hybrid reformatter}

\author[a]{Izabela Spaleniak}
\author[a]{David G. MacLachlan}
\author[b]{Itandehui Gris-S\'{a}nchez}
\author[a]{Debaditya Choudhury}
\author[c]{Robert J. Harris}
\author[a]{Alexander Arriola}
\author[c]{Jeremy R. Allington-Smith}
\author[b]{Timothy A. Birks}
\author[a]{Robert R. Thomson}
\affil[a]{SUPA, Institute of Photonics and Quantum Sciences, Heriot-Watt University, Edinburgh, EH14 4AS, UK}
\affil[b]{Department of Physics, University of Bath, Claverton Down, Bath BA2 7AY, UK}
\affil[c]{Department of Physics, University of Durham, South Road, Durham, DH1 3LE, UK}

\begin{document} 
\maketitle

\begin{abstract}
This paper reports on the modal noise characterisation of a hybrid reformatter. The device consists of a multicore-fibre photonic lantern and an ultrafast laser-inscribed slit reformatter. It operates around 1550 nm and supports 92 modes. Photonic lanterns transform a multimode signal into an array of single-mode signals, and thus combine the high coupling efficiency of multimode fibres with the diffraction-limited performance of single-mode fibres. This paper presents experimental measurements of the device point spread function properties under different coupling conditions, and its throughput behaviour at high spectral resolution. The device demonstrates excellent scrambling but its point spread function is not completely stable. Mode field diameter and mode barycentre position at the device output vary as the multicore fibre is agitated due to the fabrication imperfections.
\end{abstract}


\section{INTRODUCTION}
\label{sec:intro}

Multimode fibres have been heavily exploited in astronomy to efficiently couple the light from a telescope focus and transmit it to an astronomical instrument, such as a spectrograph. However multimode fibres suffer from an effect called \emph{modal noise} which is caused by the interference between the fibre modes and refers to noise introduced by the variation of the light pattern in the multimode fibre output \cite{Mahadevan2014}. The modal noise effect is observed when \cite{Rawson1980, Mahadevan2014}:
(i) the  source spectrum is sufficiently narrow (using a narrow bandwidth laser source or in a high-resolution spectrum) forming a speckle pattern at the fibre output, (ii) some form of spatial filtering is present at the output of the fibre, (iii) the fibre length and position are affected by movement, stress or ambient temperature variation or a change in the source wavelength or a shift in the input illumination. 

When fibres are used in high resolution spectroscopy, we encounter all of these factors: narrowband light, spatial filtering in the spectrograph by e.g. beam truncation, spatially dependent grating efficiency. Finally the optical fibre linking the telescope and the spectrograph is often under movement and affected by ambient temperature variation and stress. The input illumination also can vary during observations due to errors in the telescope pointing as well as the astronomical seeing variation. 
All of these effects cause the resulting speckle pattern to fluctuate over time and result in an unstable point spread function (PSF) of the spectrograph and variation of the signal intensity. The mode coupling is highly wavelength dependent and if we assume that conditions (ii) and (iii) are constant, then the signal intensity will vary along the wavelength with an oscillating pattern of sub-nm period\cite{Chen2006b}. As soon as the fibre or coupling into the fibre is disturbed, the periodic pattern changes. On a telescope, normally this change occurs in a timescale of $\sim$\,tens of seconds which is less than the usual time between calibration and observation, introducing uncertainties in the spectral measurements. What is more, the light from a calibration source is often delivered by a single-mode fibre to the multimode fibre link, creating a different speckle pattern to the one created by the astronomical source. 

To mitigate and average out the pattern variation fluctuation, many types of fibre scramblers have been demonstrated and are in use, e.g. fibre agitators \cite{Baudrand2001,Daino1980,Corbett2006,Lemke2011}, fibre of various geometries (e.g. square, octagonal) \cite{Avila2010},  beam homogenizers \cite{Avila2010}, ball lens scramblers \cite{Halverson2015}, integrating sphere with a moving diffuser \cite{Mahadevan2014}. However the use of single-mode fibres in high-precision Doppler spectrographs has become very attractive in recent years: not only can they provide a stable PSF and an absence of modal noise, but they also allow for the use of a smaller spectrograph with the associated cost and complexity benefits. However, a major obstacle in using single-mode fibres on seeing-limited telescopes is the inefficient coupling of light into the fibre. There has been recent work showing an efficient coupling into single-mode fibres using extreme adaptive-optics systems \cite{Jovanovic2014} but most telescopes still operate in the seeing-limited regime and exhibit poor coupling into the single-mode fibre. 

In 2005 Leon-Saval et al. \cite{Leon-Saval2005} proposed and demonstrated mode splitters, so called \emph{photonic lanterns}\cite{Leon-Saval2013},  devices which can convert a multimode signal into multiple single-mode signals and vice versa. Since then, there has been a lot of work to improve the device efficiency using either a fibre platform \cite{Noordegraaf2009,Noordegraaf2010,Birks2012} or direct-write technique \cite{Thomson2011,Spaleniak2013a}. The output of the photonic lantern is intrinsically single-mode, therefore should in principle be modal-noise free. However, there have been reports of different kinds of modal noise effects in photonic lanterns \cite{Olaya2012a,NickPasp}, indicating that it should investigated in all new types of devices.

In this paper we address the issue of modal noise in a hybrid reformatter. The device is consists of a Multicore Fibre (MCF) photonic lantern and a pseudo-slit reformatter. First we present laboratory results of the variation of the device throughput signal along the wavelength. Then we present the results of the near-field stability of the pseudo-slit.

\section{DEVICE FABRICATION AND DESCRIPTION}
\label{sec:device_fabrication}
\begin{figure} [ht]
   \begin{center}
   \begin{tabular}{c} 
   \includegraphics[width=0.95\linewidth]{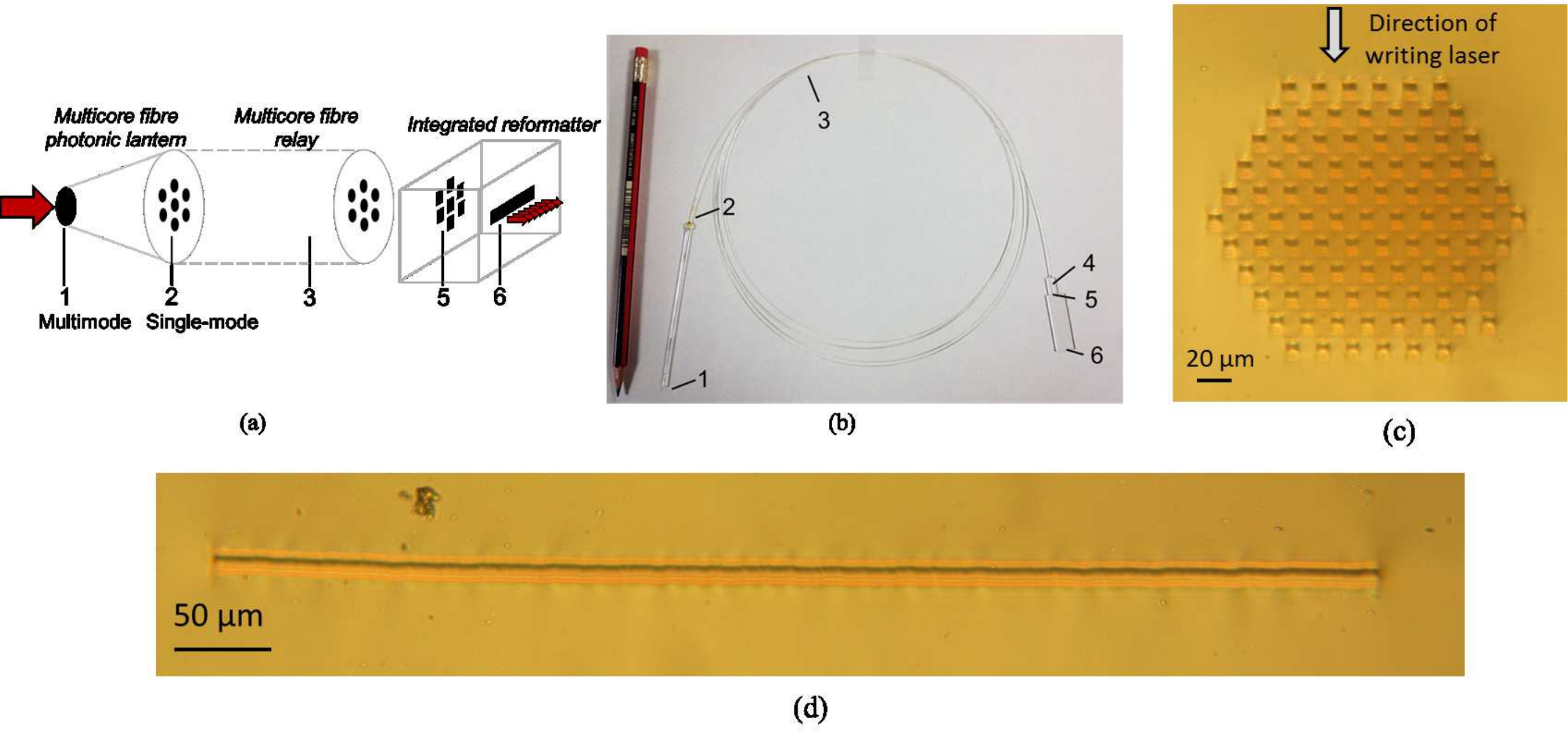}
	\end{tabular}
	\end{center}
   \caption[example] 
   { \label{fig:device_schematic} 
(a) Schematic of the simplified hybrid reformatter. (b) Image of the hybrid reformatter: (1) multimode input of the photonic lantern, (2) multicore fibre (single-mode output of the photonic lantern), (3) multicore fibre, (4) interface V-groove, (5) single-mode input of the on-chip reformatter, (6) pseudo-slit output of the reformatter. (c) Microscopic image of the single-mode end of the reformatter. (d) Microscopic image of the pseudo-slit end of the reformatter.}
   \end{figure}

The hybrid device consists of a multicore fibre (MCF) photonic lantern with a few metres of MCF and an integrated slit reformatter (Fig.\,\ref{fig:device_schematic}). 

The device comprises 92 cores: 91 are arranged in hexagonal pattern and 1 extra waveguide is used for device orientation (Fig.\,\ref{fig:device_schematic}(c)). The multimode input has a diameter of $\sim$\,43\,$\upmu$m. The single-mode cores in MCF of 9\,$\upmu$m in diameter are separated by 17.6\,$\upmu$m, which means that there is a significant coupling between the cores. This arrangement was matched by the individual cores in the single-mode of the reformatter (Fig.\,\ref{fig:device_schematic}(a)) with 6.2\,$\upmu$m-wide waveguides. Despite the difference in the MCF core diameter and waveguide diameter their MFDs are closely matched. The integrated reformatter rearranges a 2D array of waveguides into a 1-D continuous sequence of waveguides forming a planar waveguide of 570\,$\upmu$m in width, a so-called pseudo-slit. The resulting pseudo-slit was designed to be single-mode across but multimode along its length. The main advantage of creating a continuous pseudo-slit instead of discrete single-mode waveguides at the device output, is that it can in principle reduce the required detector size and cross-dispersion when used in a spectrograph.

The multimode end of the photonic lantern was fabricated by inserting the multicore fibre into a low refractive index jacket and tapering it down. Waveguides are tapered until they no longer can efficiently guide the light. The low refractive index jacket forms the cladding of the multimode fibre. The integrated reformatter was fabricated using ultrafast laser inscription, where femtosecond laser pulses are focused into a glass substrate and modify its refractive index. Light from a fibre laser system (MenloSystems, BlueCut) of 1030\,nm wavelength, 500\,kHz repetition rate and 350\,fs long pulses was focused into an alkali-free borosilicate glass substrate (Schott AF45) with a 0.4\,NA aspheric lens. The sample was mounted on high-precision x-y-z air-bearing translation stages (Aerotech) allowing to create 3D optical circuits. Each of the 92 and 6.2$\upmu$m  wide waveguides was formed using a multiscan technique \cite{Said2004} where each of 31 scans through the laser focus created a  0.2\,$\upmu$m wide refractive index modification. As can be seen from (Fig.\,\ref{fig:device_schematic}(c,d)) the refractive index modification along the direction of writing laser consists of a top short bright area, middle dark area and bottom long bright area, which is fairly typical for AF45 glass. Despite the complicated refractive index profile, a Gaussian mode propagates along the single mode waveguides.

\section{LABORATORY MEASUREMENTS}
In this section we describe two types of measurements done to assess the modal noise issue in the hybrid reformatter. The first measurement looks at the throughput changes with wavelength. The second one looks at the stability of the near-field of the pseudo-slit.

\subsection{Wavelength Dependence of the Throughput}
\label{sec:title}

   \begin{figure} [ht]
   \begin{center}
   \begin{tabular}{c} 
   \includegraphics[width=0.7\linewidth]{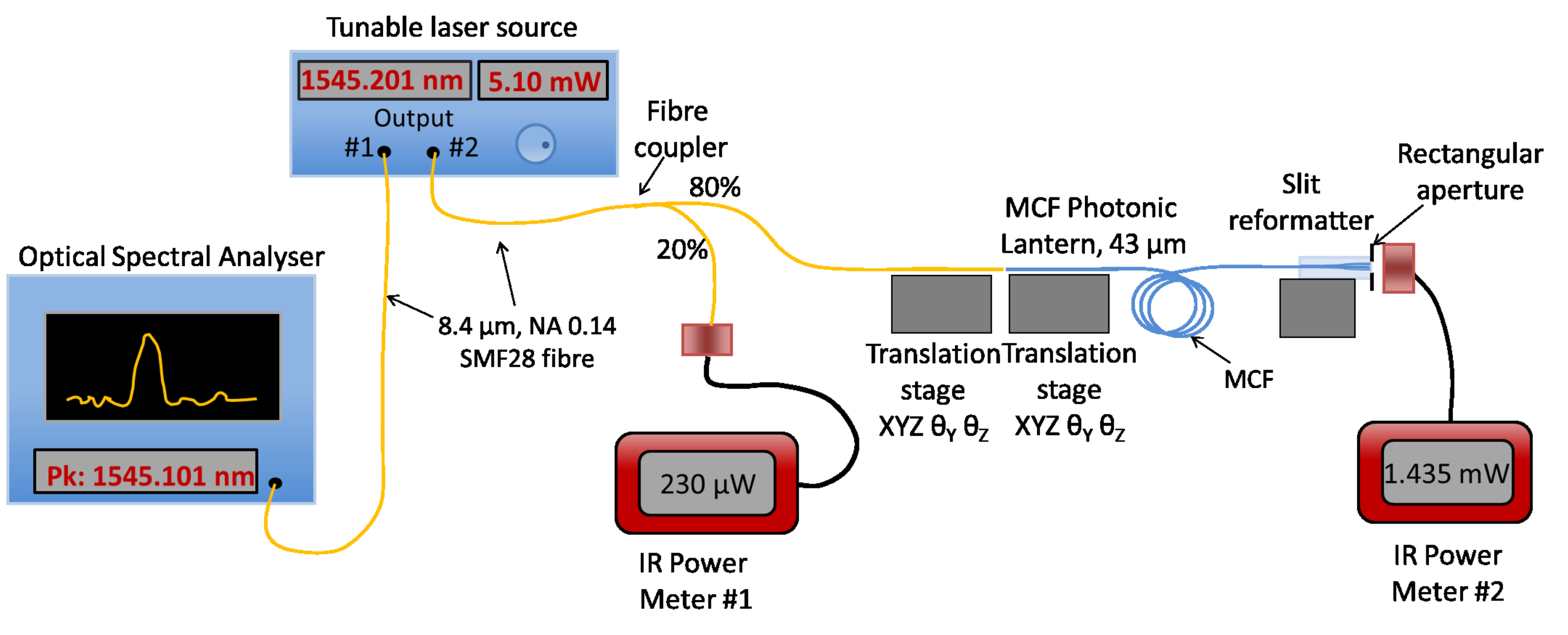}
	\end{tabular}
	\end{center}
   \caption[example] 
   { \label{fig:Modal_noise_schematic_wavelength} 
Schematic of the laboratory setup to measure the throughput of the device as a function of wavelength.}
   \end{figure} 

One of the manifestations of modal noise is a periodic device throughput variation with wavelength at high spectral resolution. The best way to assess the modal noise is to build a spectrograph but this is outwith the scope of this paper. As a "close enough'' approximation we used a narrow bandwidth tunable laser (Anritsu MG9638A) and measured the device throughput variation across wavelength range (Fig.\,\ref{fig:Modal_noise_schematic_wavelength}) at high spectral resolution. The laser source has two outputs, allowing for a simultaneous measurement of the output wavelength using an optical wavelength analyser. A 20/80 fibre coupler was used to monitor the laser power. The light was delivered by a single-mode fibre (SMF28) into the multimode end of the photonic lantern. The signal at the slit output of the lantern was measured using a power meter. To ensure that only guided light was measured on power meter \#2, a rectangular aperture with dimensions $\sim$20\,$\upmu$m larger than the pseudo-slit was applied. The resulting output power was normalised to the reference power measured at power meter \#1 and corrected for the spectral characteristics of the fibre coupler.

\begin{figure}[!htpb]
\centering
\subfloat[\label{fig:modal_noise_coarse}]{\includegraphics[width=0.45\linewidth]{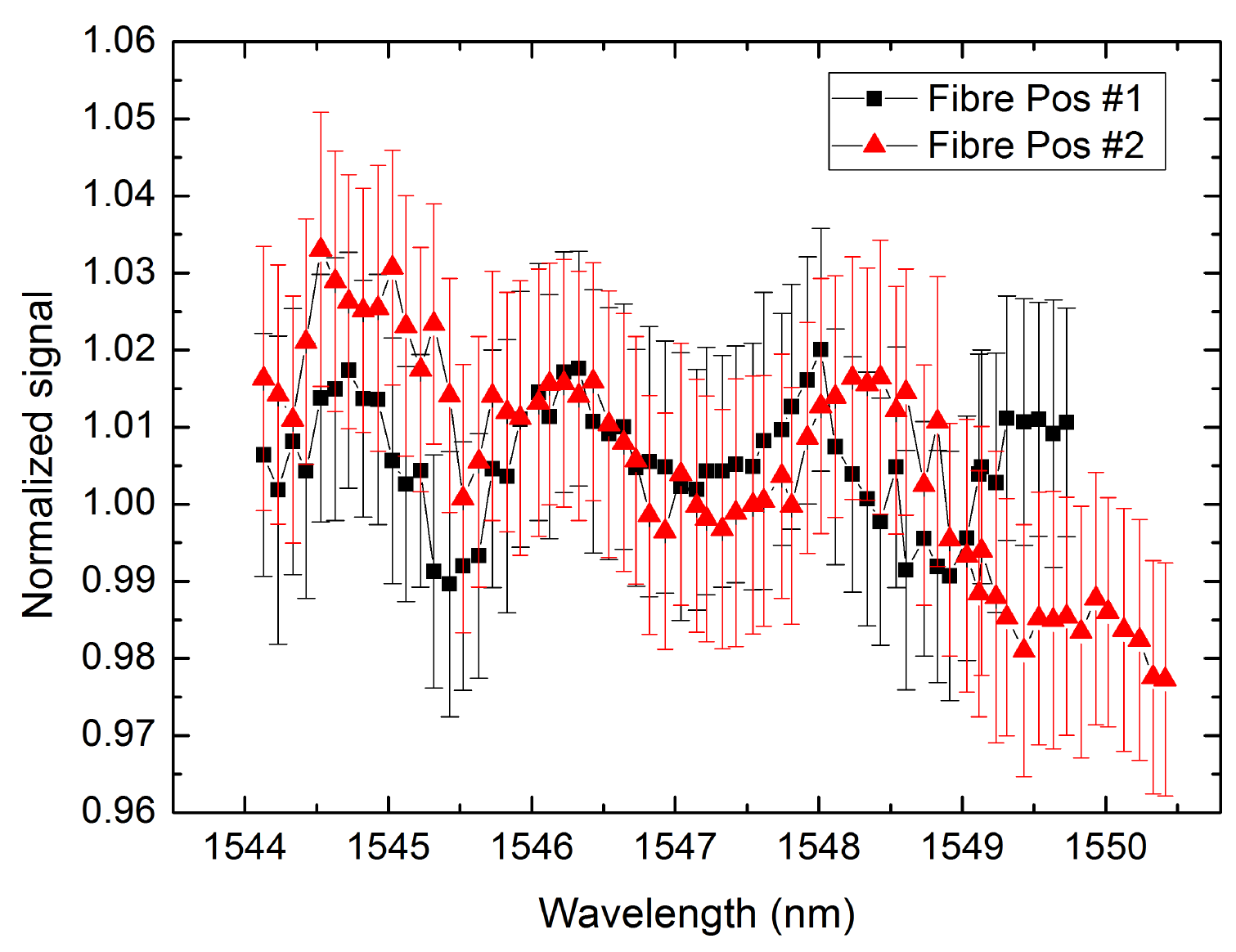}}
\subfloat[\label{fig:modal_noise_fine}]{\includegraphics[width=0.45\linewidth]{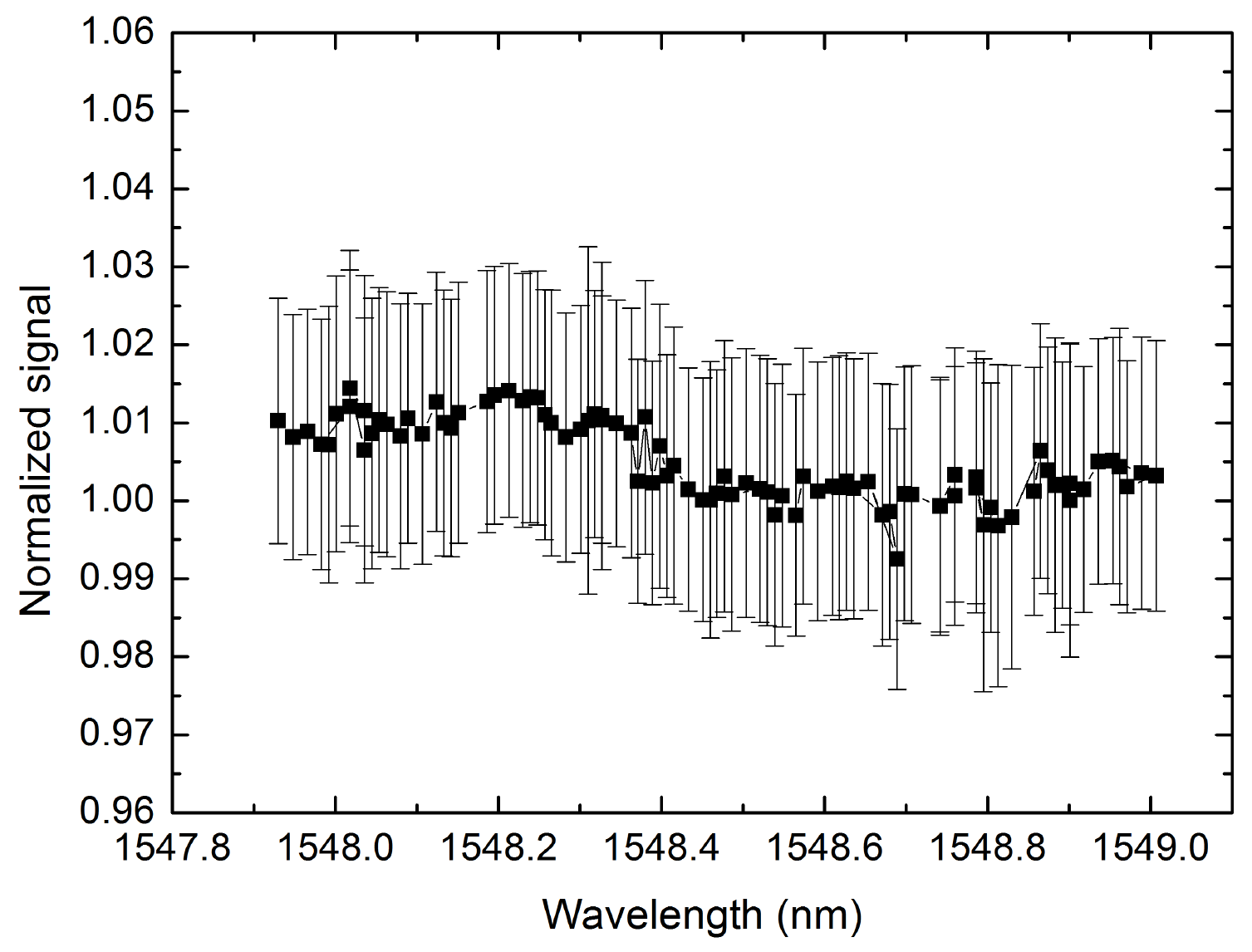}}
\caption{Throughput of the hybrid reformatter as a function of wavelength for (a) low and (b) high spectral resolution. The wavelength increments were $\sim$\,0.1\,nm ($R$\,$\sim$\,15,500) and $\sim$\,0.01\,nm ($R$\,$\sim$\,155,000) respectively. The low resolution measurement was repeated for two different positions of the SM fibre input producing similar results.}
\label{fig:modal_noise}
\end{figure}

Two types of measurements were done imitating the ``low'' and ``high'' spectral resolution. The low resolution covered 1544\,nm\,-\,1550\,nm wavelength range with 0.1\,nm increments, which corresponds to $R$\,$\sim$\,15,500, while the high resolution covered 1548\,nm\,-\,1549\,nm region with 0.01\,nm increment, which corresponds to $R$\,$\sim$\,155,000. Figure\,\ref{fig:modal_noise} shows the resulting normalised throughput variation over the wavelength. As can be seen the throughput was measured to be insensitive to the wavelength variation at both ``low'' and ``high'' resolution. Small throughput variation is within the measurement error. The uncertainty of $\sim$\,3.5\% (marked as error bars in Fig.\,\ref{fig:modal_noise}) comes from the polarisation dependent losses. Hence, we can conclude that we do not observe the device throughput variation with wavelength.

\subsection{Pseudo-slit Stability}
   \begin{figure} [ht]
   \begin{center}
   \begin{tabular}{c} 
   \includegraphics[width=0.7\linewidth]{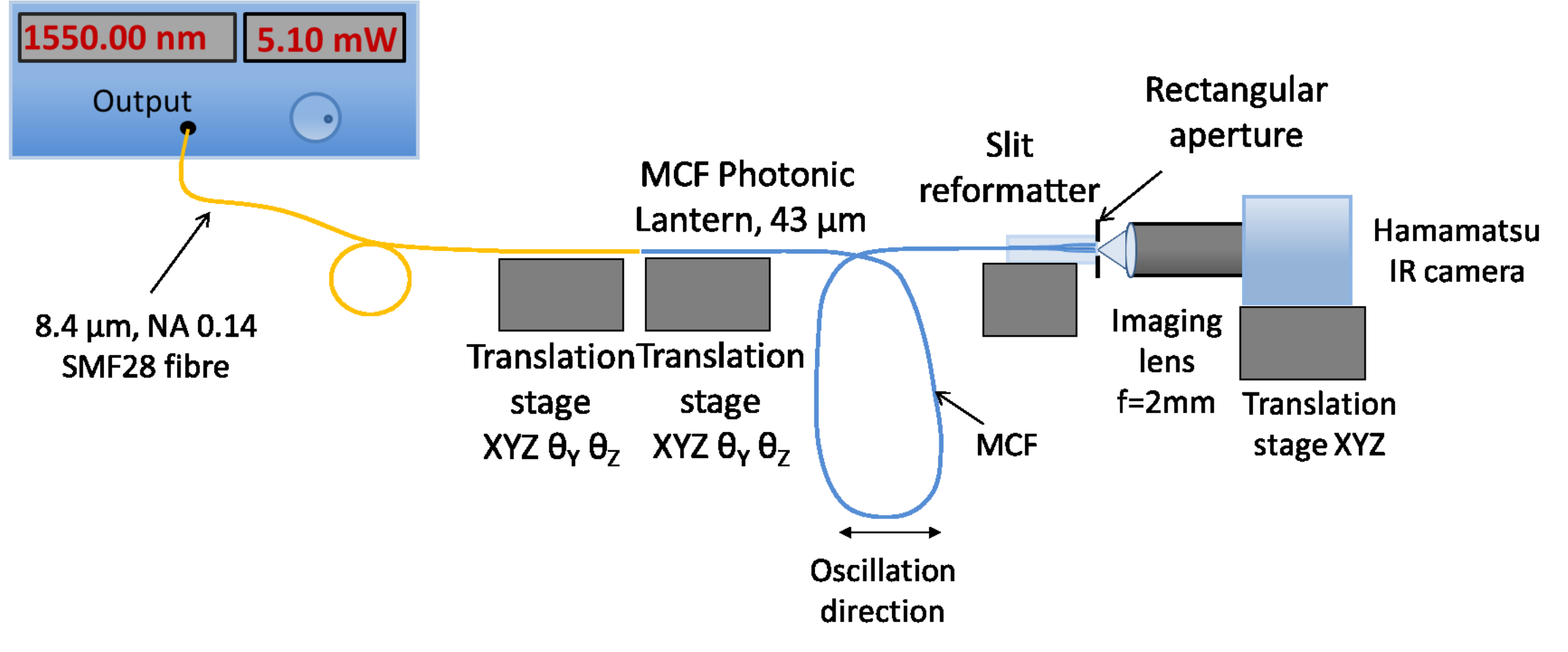}
	\end{tabular}
	\end{center}
   \caption[example] 
   { \label{fig:Modal_noise_schematic_oscillation} 
Schematic of the laboratory setup to measure the MFD variation of the pseudo-slit as the MCF is moved in oscillating manner.}
   \end{figure} 

In this experiment we inspected the stability of the device output. Single-mode light injected into the multimode end of the photonic lantern produces a speckle pattern at the pseudo-slit end, which changes as the coupling changes or the MCF moves. As described in Sec.\,\ref{sec:device_fabrication} cores in the MCF are coupled therefore any disturbance of the MCF changes the coupling between the modes resulting in a different speckle pattern at the pseudo-slit. To assess the stability of the pseudo-slit end, we measured the variation of the mode field diameter (MFD) and its barycentre (central position of the mode).

Figure\,\ref{fig:Modal_noise_schematic_oscillation} shows the schematic of the measurement setup. Light from a 1550\,nm laser is coupled into a single-mode fibre (Corning SMF28) and injected at approximately the centre of the multimode end of the MCF photonic lantern. A length of $\sim$\,2\,m of the MCF is hanging off the table and moved by hand in an oscillating manner. Pseudo-slit output is imaged onto an IR detector (Hamamatsu  C10633) using an AR coated aspheric lens (Thorlabs C150TME-C). As described in Sec.\,\ref{sec:device_fabrication} the pseudo-slit is long and thin, so in order to achieve high enough resolution across the pseudo-slit to accurately probe its properties, only a $\sim$\,76\,$\upmu$m-long part of along the slit was imaged. A set of 800 frames of 2\,ms-long exposures was recorded and analysed. The actual acquisition time was $\sim$\,5\,s due to the read-out time and data transfer delay. The measurement was repeated at left and right (centre$-20\,\upmu$m and centre$+20\,\upmu$m, respectively) positions across the multimode end of MCF photonic lantern.

\begin{figure}[!htpb]
\centering
\includegraphics[width=0.8\linewidth]{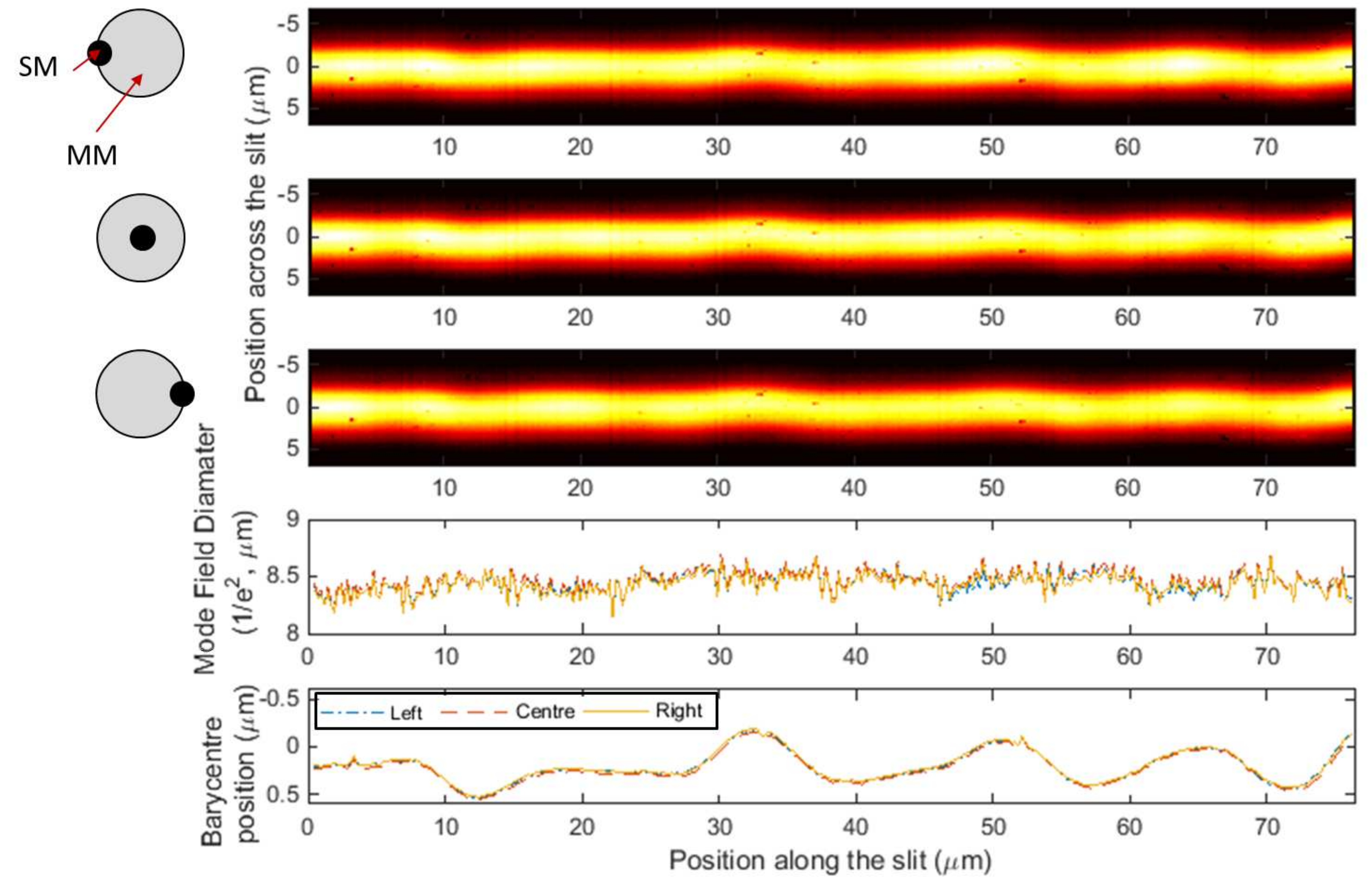}
\caption{Top panels: Near-field images of a $\sim$\,76\,$\upmu$m long representative part of the slit, when coupling SM light into the multimode end of the MCF photonic lantern at three different input positions represented on the left of each image. The images are a result of recording and stacking 800 frames while agitating the MCF. Bottom panels: mode field diameter and position of the barycentre calculated along the slit.}
\label{fig:Slits_images_barycentre}
\end{figure}

In the first analysis all 800 frames for each injection position were stacked. A Gaussian function was fitted to each column of pixels and MFD (measured at 1/e$^2$) and barycentre (position of the centre of the fit) were compared. Figure~\ref{fig:Slits_images_barycentre} shows the results. The top three panels show the slit images obtained by stacking 800 frames with different speckle patterns for three different injection positions. The two bottom panels show the mode field diameter and barycentre calculated along the pseudo-slit for each pixel. We can see that the shift of the  MFD and barycentre due to the shift of the injection positions is  comparatively small. The difference in MFD was measured to be under 0.12\,$\upmu$m ($\sim$\,1.5\,\%) and the shift in the barycentre was position under 0.07\,$\upmu$m ($\sim$\,0.8\,\% of the MFD). However it is clear that the slit is neither straight nor uniform in its width,  with variations of up to 0.5\,$\upmu$m ($\sim$\,6\,\%) in the measured mode field diameter and 0.7\,$\upmu$m ($\sim$\,8\,\% of MFD) in the barycentre position along this section of the pseudo-slit.

\begin{figure}[!htpb]
\centering
\includegraphics[width=0.45\linewidth]{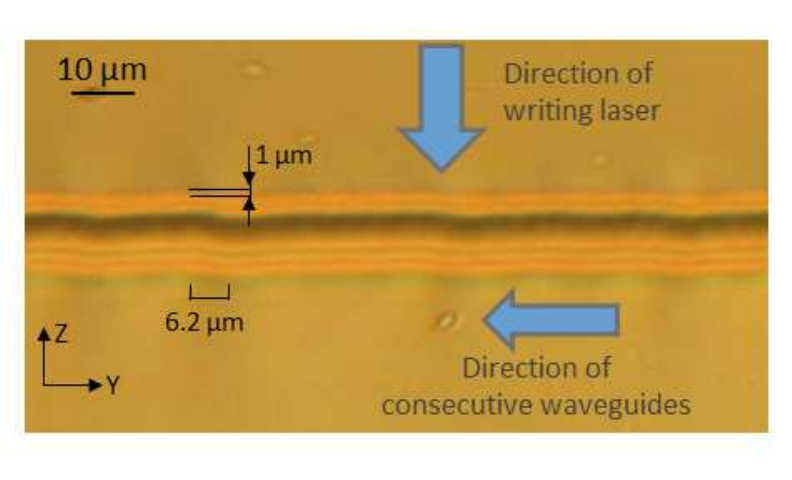}
\caption{Micrograph of the pseudo-slit section showing its uneven profile.}
\label{fig:slit_micro}
\end{figure}

The non-straight slit (Fig.\,\ref{fig:slit_micro}) is a result of the reformatter fabrication setup and process. Due to the laboratory environment instability, the top of the sample had to be refocused  before writing each of the 31 scans to form a waveguide. However, a lens used to focus the femtosecond laser beam in the sample had NA of 0.4, producing a relatively large depth of focus of $\sim$\,4\,$\upmu$m. As a result our current fabrication and alignment setup allows to focus the beam only within 1\,$\upmu$m. 

\begin{figure}[!ht]
\centering
\includegraphics[width=0.7\linewidth]{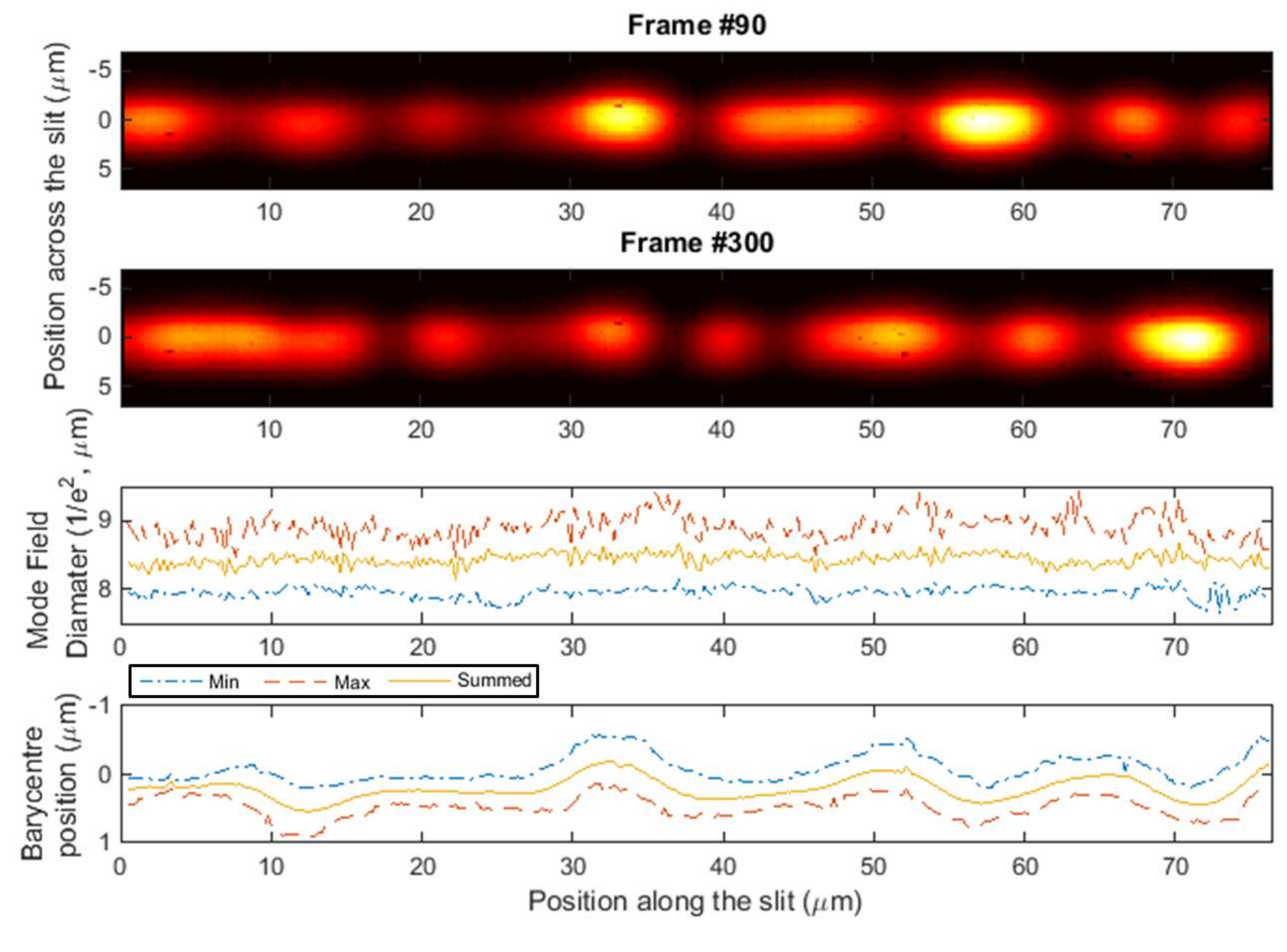}
\caption{Top panels: Near-field images of  $\sim$\,76\,$\upmu$m long representative part of the slit, when coupling SM light into the MM end of the MCF  lantern at one (left) input position (as in Fig.\,\ref{fig:Slits_images_barycentre}). The images are examples of the speckle distribution in one frame while agitating the MCF. Bottom panels: maximum, minimum and summed MFD and position of the barycentre calculated along the slit.}
\label{fig:Slits_images_non_agitated}
\end{figure}

In the second analysis of the acquired data, 800 frames with different excited modes and speckles distribution were compared for one fixed light injection position. A Gaussian function was fitted to each column of pixels (meaning consecutive positions along the pseudo-slit) of each frame. A maximum and minimum of MFD and maximum and minimum of barycentre position for each column of pixels were found amongst the various speckle patterns. Top two panels in Fig.\,\ref{fig:Slits_images_non_agitated} show two examples of speckle patterns. The two bottom panels show the minimum, maximum and summed  (result of stacking up all the frames as in Fig.\,\ref{fig:Slits_images_barycentre}) values of MFD and barycentre.  It is apparent that both MFD and barycentre variation is significant, with the difference between maximum and minimum of MFD being 1\,$\upmu$m (12\,\%) on average and the difference between maximum and minimum of barycentre being 0.6\,$\upmu$m (7\,\% of the MFD) on average. It can be noticed from the barycentre plot that the difference between the maximum and minimum is smaller ($\sim$0.3\,$\upmu$m) along the straighter sections of the pseudo-slit (at 5\,$\upmu$m and 20\,$\upmu$m) and larger ($\sim$0.8\,$\upmu$m) around the large offsets (at 10\,$\upmu$m and 32\,$\upmu$m). 

The variation of MFD and barycentre are of course undesirable in high-precision Doppler spectrographs. Any instability of PSF will introduce noise and uncertainties in the radial velocity measurement. 

\subsection{Pseudo-slit Stability Simulation}

To assess the impact of the pseudo-slit non-straightness on modes along the slit, BeamProp (package of RSoft\cite{RSoft}) was used to simulate the pseudo-slit and solve the propagating modes. We created a slab waveguide composed of 82 rectangular waveguides, which were 1.55\,$\upmu$m in width and 6.2\,$\upmu$m in height rectangular waveguides parallel to each other which are offset by up to $\pm 0.4$\,$\upmu$m following the trend of the barycentre position from Fig.\,\ref{fig:Slits_images_barycentre}. The length of the waveguides was set to 50\,mm allowing all the mode solutions to converge. As a reference, a straight, uniform slab waveguide matching the size of the non-straight waveguide slab was also created. 

\begin{figure}[!htpb]
\centering
\includegraphics[width=0.7\linewidth]{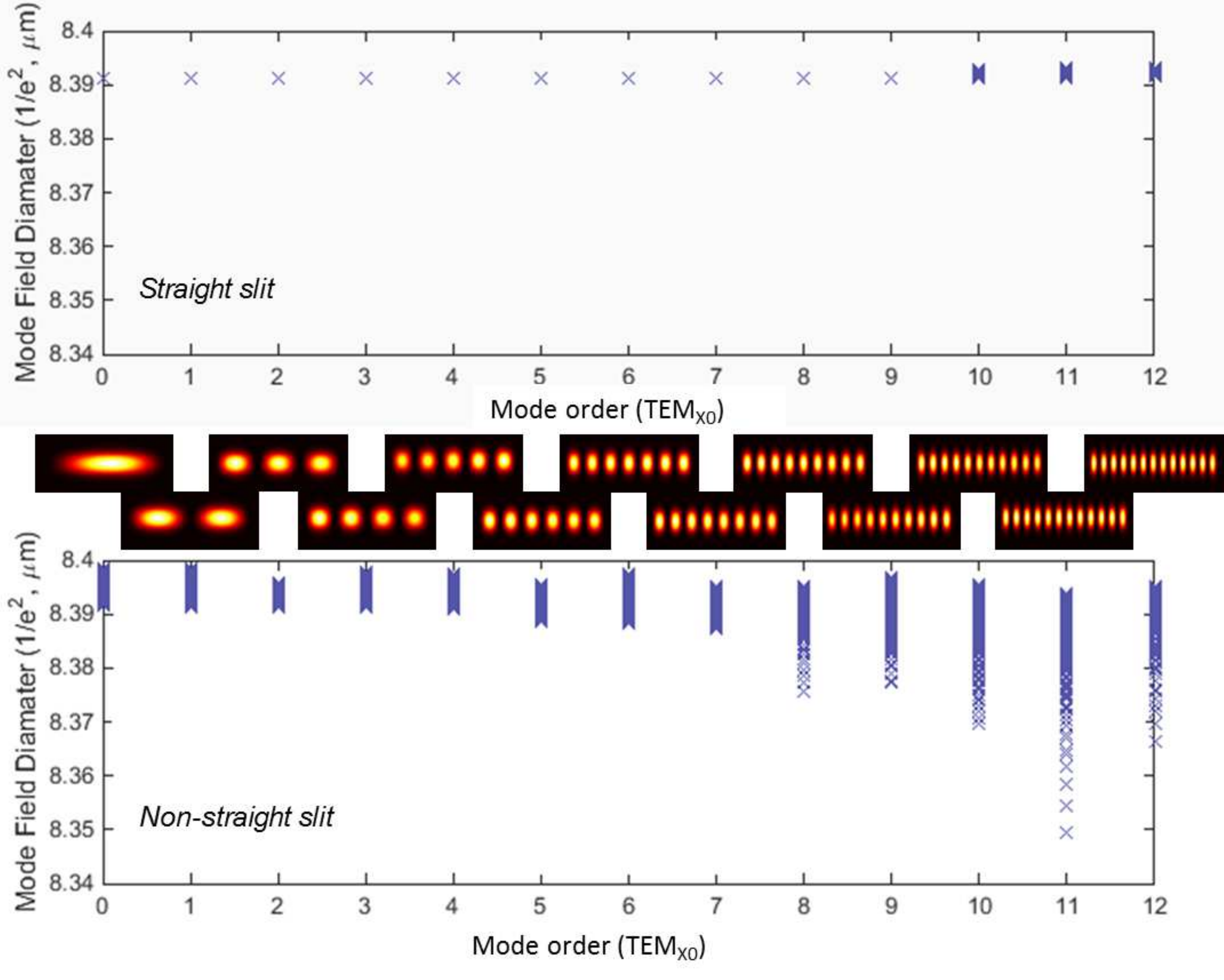}
\caption{Comparison of the MFD range along the simulated slit for a straight and non-straight slab waveguide.}
\label{fig:MFD_and_modes_simulation}
\end{figure}

Both simulated waveguides support 12 TEM modes, as depicted  in Fig.\,\ref{fig:MFD_and_modes_simulation}. To analyse the data, a Gaussian function was fitted across the the slit for each mode and MFD and barycentre position were calculated. Figures~\ref{fig:MFD_and_modes_simulation} and \ref{fig:barycentre_and_modes_simulation} show combined values of MFD and barycentre position, respectively, for each of the modes calculated along the waveguide length. It can be seen from Fig.\,\ref{fig:MFD_and_modes_simulation} that for the non-straight slit the spread of MFDs increases for higher order modes, whereas it stays constant for the straight slit.  It can be deducted that the higher order modes, which are smaller, are much more affected by small changes in waveguides geometry. From Fig.\,{\ref{fig:barycentre_and_modes_simulation} we can again see that each of the modes in the non-straight slit has a much higher variation in the barycentre position of the modes, compared to the straight slit.  

Intuitively we can think about this problem as follows: by measuring the MFD and barycentre position we look at the projection of the modes/speckles. With different angles of various speckles we see that the projection angle changes and so do the measured MFD and barycentre position.

\begin{figure}[!htpb]
\centering
\includegraphics[width=0.8\linewidth]{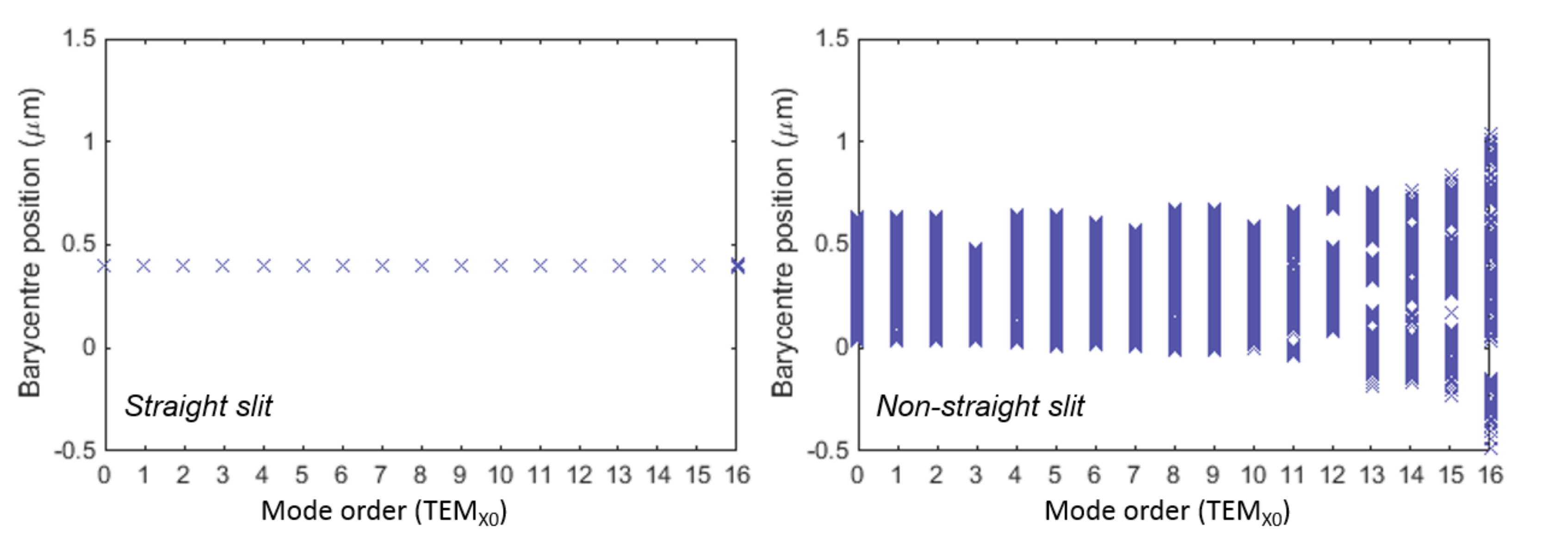}
\caption{Comparison of the barycentre range along the simulated slit for a straight and non-straight slab waveguide.}
\label{fig:barycentre_and_modes_simulation}
\end{figure}

We can see from the simulations that the non-straight pseudo-slit is responsible for the instability of the PSF of the pseudo-slit in the hybrid reformatter. This effect is differs from the modal noise in mutlimode fibres but nevertheless will contribute to radial velocity measurement errors. We can suggest two possible ways to address this. The first is to improve the laboratory stability and inscription setup precision to fabricate ``perfectly'' straight  pseudo-slit. The second and somewhat easier solution is to produce a discrete, non-coupled set of waveguides at the device output. However, this solution will require waveguides to be separated by about $\sim$20-30\,$\upmu$m. Larger separation between the waveguides will require larger detector area and higher cross-dispersion compared to the continuous pseudo-slit when used in a spectrograph.

\section{CONCLUSIONS}

Modal noise in multimode fibres is a limiting aspect of high-resolution Doppler spectroscopy. Photonic lanterns, in principle do not exhibit this effect, however previous studies have shown that it not always the case. In this paper, we performed two types of measurements to address the modal noise effect in a hybrid device -- a component consisting of a MCF photonic lantern and direct-write pseudo-slit reformatter. 

We did not find any significant and periodic variation of the device throughput over wavelength at high spectral resolution, which means that we did not detect the modal noise effect in the device. We also found that the device is a very good mode scrambler and gentle agitation of MCF produces a homogeneous image at the device output. However we also found that the PSF of the device output is not completely stable, so it varies depending on excited modes and produced speckles. We showed that this variation originates from the fact that the pseudo-slit is not perfectly straight along its length. 

The observed PSF instability can contribute to errors in the radial velocities measurements. To overcome this problem we propose either improvement of the laboratory environment and inscription setup stability to produce a ``perfectly'' straight pseudo-slit or fabrication of a one dimensional array of discrete single-mode waveguides at the device output.

\acknowledgments 
 
R.R.T. gratefully acknowledges funding from the STFC in the form of an STFC Advanced Fellowship (ST/H005595/1) and also from Renishaw plc. R.R.T and T.A.B acknowledge funding via the STFC-PRD scheme (ST/K00235X/1). R.R.T, T.A.B and J.R.A-S thank the European Union for funding via the OPTICON Research Infrastructure for Optical/IR Astronomy (EU-FP7 226604). D.G.M. is supported by an EPSRC Ph.D studentship. R.J.H. and J.R.A-S. gratefully acknowledge support from the Science and Technology Facilities Council (STFC) in the form of a Ph.D studentship (ST/I505656/1) and grant (ST/K000861/1).

\bibliography{spie2016_references} 
\bibliographystyle{spiebib} 

\end{document}